\documentclass[aps,pre,twocolumn,preprintnumbers,amsmath,amssymb,nofootinbib]{revtex4-2}

\usepackage{epsfig}
\usepackage{epstopdf}
\usepackage{lipsum}
\usepackage{multirow}
\usepackage{boldline}
\usepackage{xcolor}

\begin{document}
	\title{Asymmetry of social interactions and its role in link predictability:\\the case of coauthorship networks}

	\author{Kamil P. Orzechowski, Maciej J. Mrowinski, Agata Fronczak and Piotr Fronczak}
	\affiliation{Faculty of Physics, Warsaw University of Technology,
		Koszykowa 75, PL-00-662 Warsaw, Poland}
	\date{\today}
	
	\begin{abstract}
		The paper provides important insights into understanding the factors that influence tie strength in social networks. Using local network measures that take into account asymmetry of social interactions we show that the observed tie strength is a kind of compromise, which depends on the relative strength of the tie as seen from its both ends. This statement is supported by the Granovetter-like, strongly positive weight-topology correlations, in the form of a power-law relationship between the asymmetric tie strength and asymmetric neighbourhood overlap, observed in three different real coauthorship networks and in a synthetic model of scientific collaboration. This observation is juxtaposed against the current misconception that coauthorship networks, being the proxy of scientific collaboration networks, contradict the Granovetter's strength of weak ties hypothesis, and the reasons for this misconception are explained. Finally, by testing various link similarity scores, it is shown that taking into account the asymmetry of social ties can remarkably increase the efficiency of link prediction methods. The perspective outlined also allows us to comment on the surprisingly high performance of the resource allocation index -- one of the most recognizable and effective local similarity scores -- which can be rationalized by the strong triadic closure property, assuming that the property takes into account the asymmetry of social ties.
	\end{abstract}
	\maketitle

	\section{Introduction}
	
Social networks representing patterns of human interactions have been the subject of both empirical and theoretical research since at least the middle of the last century \cite{1994bookWasserman, 2010bookEasley}. And although, over the past two decades, due to the rise of the Internet followed by the increased availability of large datasets on human interactions, methods of social network analysis have changed a lot, the basic challenge behind these analyses remained the same: To understand human behaviour \cite{2020ScienceLazer, 2012manifestoSNA, 2012NatureGiles}. In particular, questions that keep recurring in the literature of the field are: How, depending on the context studied, people establish social ties? To what extent can the evolution of a social network be modelled using features intrinsic to the network itself? Is it possible to predict existing but undisclosed or intentionally hidden connections based on those recorded? Finally, what influences the strength of social ties, and can these strengths be inferred from the binary link structure?

In what follows, building upon results of our previous paper \cite{2022SciRepFronczak}, we refer to last two of the above questions. We show that in order to better understand weight-topology correlations in social networks, it is necessary to use measures that formally take into account asymmetry of social interactions, which may arise, for example, from differences in ego-networks of connected nodes. A simple argument in favour of this statement can be drawn from the theory of complex networks (more specifically, from the degree-based mean field approach \cite{HMF1, HMF2, HMF3}). In particular, in social networks with fat tailed node degree distributions the sizes of ego-networks of two connected nodes may differ considerably. This means that their common neighbours can be a significant part of the neighbourhood of one node and an insignificant part of the neighbourhood of the other, resulting in a completely different perception of the size of the common neighbourhood on both ends of the connection. This indicates that the observed \textit{absolute} tie strength is a kind of compromise, which depends on the \textit{relative} strength of the tie as seen from its both ends.  

Recently, similar findings have been made in Ref.~\cite{2018SciRepMattie}, where the concept of the \textit{social bow tie} has been introduced. Bow tie consists of a focal tie and all nodes connected to either or both of the two focal nodes. In the mentioned study, a number of topological metrics quantifying properties of such a bow tie (including sum and absolute difference of clustering coefficients of connected nodes) have been investigated through machine learning and regression models in two different types of social networks (e.g. call network of mobile phone users). The main conclusion from this study was that in the considered networks tie strength depends not only on the properties of shared friends but also on those tied to only one person, hence introducing a fundamental asymmetry to social interaction. Despite interesting conclusions, the authors however failed to identify the most predictive, quantitative indicators of tie strength, basing their findings on a broad spectrum of different structural properties of bow ties. From this perspective, in the face of the growing interest in measuring and predicting social ties (see e.g.~\cite{2015PNASLu, 2019JASISTKim, 2021JASISTZou, 2021ComPhysUbaldi}), an important step towards finding such an informative metric has been made in our recent paper on Granovetter's theory in coauthorship networks \cite{2022SciRepFronczak}.

Historically, the Granovetter's theory \cite{1973Granovetter,1995bookGranovetter} is of importance to weight-topology correlations in social networks, as Mark Granovetter was the first to distinguish between strong and weak social ties. He treated ties as if they were positive and symmetric, and suggested that, from a network structure perspective, tie strength between any two people should increase with the number of their mutual friends. In line with this hypothesis, several intuitive network measures, such as the neighbourhood overlap \cite{2007PNASOnnela}, have been proposed to characterize the aforementioned correlations. Unfortunately, contrary to expectations, performance of these indicators turned out to be not very satisfactory: sometimes confirming \cite{2007NewJPhysOnnela, 2010SocNetSzell, 2012RoyalSuvakov}, and sometimes saying nothing \cite{2012NatPhysPajevic, 2021ComPhysUbaldi}, or even contradicting \cite{2012EPLPan, 2014PREKe} the Granovetter's hypothesis. 

And while no systematic attempts have been made to explain the poor performance of these indicators to date, recent studies \cite{2020EPJDataSaramaki, 2018SciencePark, 2022SciRepFronczak} may point to some reasons of their failure. For example, in Ref.~\cite{2020EPJDataSaramaki}, by analysing a large mobile-phone dataset, it has been shown that temporal features of social ties (such as the number of days with calls, number of bursty cascades, typical time of contacts, etc.) are related to both their strength and topological features of their nearest network neighbourhood. In Ref.~\cite{2018SciencePark}, analysis of population-scale mobile-telephone and Twitter data has revealed that unembedded long-range connections (i.e. with no nearest neighbours and long second-nearest paths) can be as strong as embedded ones (with non-zero neighbourhood overlap). Finally, in Ref.~\cite{2022SciRepFronczak}, using a large scale real coauthorship network, we have provided evidence that the key to understand weight-topology correlations in social networks is to reject the assumption of the symmetry of social ties that is commonly used in scientific research.

It is no wonder then that such indicators as the number of common neighbours or neighbourhood overlap, when used in link prediction methods, gave results comparable (and often even worse) to the typical measures of nodes' similarity \cite{2011PhysALu, 2017ACMMartinez}, such as: the Adamic-Adar index \cite{2003SocNetwAdamic} or the resource allocation index \cite{2009EPJBZhou}. In fact, the above-mentioned problems are particularly evident in coauthorship networks, in which many independent studies \cite{2012EPLPan, 2012NatPhysPajevic, 2014PREKe, 2021ComPhysUbaldi, 2022SciRepFronczak} have confirmed non-monotonic (instead of strictly growing), U-shaped relation between tie strength and neighbourhood overlap of adjacent nodes that is contrary to the Granovetter's hypothesis. In our last paper \cite{2022SciRepFronczak}, using DBLP computer science bibliography database, we identified the source of this problem, pointing to inappropriate (i.e. symmetric instead of asymmetric) quantities used to study the weight-topology correlations. We have introduced new measures: asymmetric neighbourhood overlap and asymmetric tie strength which allowed the successful verification of the Granovetter's theory, and which - we believe - may be helpful in developing new link prediction methods in social networks.

In this paper, to reinforce the message of our recent contribution \cite{2022SciRepFronczak}, we investigate the weight-topology correlations in two more real coauthorship networks and in a synthetic model of scientific collaboration, which reproduces many of the properties of these networks. The motivation behind this study is twofold. First, the analyses with the use of different real data and synthetic networks are intended to validate our findings on the role of asymmetry in social ties, that were originally derived from analysis of just one dataset \cite{2022SciRepFronczak}; such validation is an important element of the research, as it shows that the results described in our previous paper are not an artefact resulting from the specificity of the only dataset used. Second, with this contribution, we would like to point out potential applications of the new network measures we introduced in \cite{2022SciRepFronczak} to the problem of link prediction in social networks; since most of the known link-prediction methods use symmetric network measures \cite{2011PhysALu, 2015PNASLu, 2017ACMMartinez}, contributions like this one are important because they increase the awareness of society that redefining traditional measures to account for link asymmetry can significantly improve their performance. 

At this point, we would like to highlight the difference between our contribution and existing research on link prediction in directed networks \cite{2014Schall, 2019Butun, 2015Wang, 2020Liu, 2013Zhang}. We are dealing here with undirected networks. Howerer, despite the lack of link directions, we exploit a natural asymmetry in studied networks that can be used to predict links more effectively. This approach is completely new.

The reminder of this paper is organized as follows. In Section~\ref{part1}, we study weight-topology correlations in three different real coauthorship networks and in a synthetic model of scientific collaboration. For this purpose, we use a new metric of local edge clustering - the asymmetric neighbourhood overlap, which extracts information about the asymmetry of social ties. In Section~\ref{part2}, we provide an in-depth discussion of different similarity scores used in classical methods of link and weight prediction in complex networks. Understanding why some of these measures are successful allows us to design new, inherently asymmetric indices that outperform existing ones. Section \ref{theend} draws conclusions of the paper.

\section{Asymmetry-based weight-topology correlations}\label{part1}

\subsection{Methods}

\subsubsection{Scientific collaboration networks}

Coauthorship networks, with nodes representing all scientists in a particular discipline and edges joining pairs who have coauthored articles \cite{2016JoIXie}, are widely accepted as proxies of scientific collaboration networks \cite{2001PNASNewman, 2004PNASNewman}. Accordingly, their properties are often compared to other proxies of social networks, such as mobile phone networks \cite{2007PNASOnnela, 2007NewJPhysOnnela, 2021ComPhysUbaldi}. In this respect, when considered as binary networks - without any additional features assigned to nodes and connections - all these networks show numerous structural similarities (e.g. high clustering, small-world effect, and skewed degree distribution \cite{2001PREaNewman}). However, when the edges are assigned weights representing, depending on the network, the number of joint publications or the number of phone calls made then, although macroscopic features of these networks (such as distributions of node strengths and edge weights \cite{2001PREbNewman}) may still be similar, their weight-topology correlations arising from the localization of strong and weak ties \textit{seem to be} completely different \cite{2012EPLPan}.

Indeed, it is widely believed that coauthorship networks show atypical weight-topology correlations compared to other - let's say \textit{typical} - social networks. Here, the term \textit{typical} refers to networks that satisfy the Granovetter's hypothesis \cite{1973Granovetter}, according to which strong social ties are associated with densely connected groups of individuals, while weaker ties act as bridges between these groups. In what follows, we take a closer look at these issues. We show that the phrase we just used, namely: \textit{seem to be} instead of \textit{are}, is not accidental, because in fact coauthorship networks show weight-topological correlations \textit{typical} of other social networks, provided that the measures used to analyse them are properly adapted to their structure. 

\subsubsection{Datasets used}

We analyse coauthorship networks built from three scientific databases containing publication records in the field of computer science and physics: the DBLP Computer Science Bibliography, the American Physical Society (APS) journal articles, and the Condensed Matter (CondMat) section of the preprint server ArXiv. In detail:

\begin{itemize}
\item DBLP is a digital library of article records published in computer science \cite{2008Tang}. In this study, we use the 12th version of the dataset (DBLP-Citation-network V12; released in April 2020 \cite{DBLP}), which contains information on approximately 4.9~M articles published mostly during the last 20 years. We ourselves processed the raw DBLP data into the form of coauthorship network and, following previous studies of similar data, we focused on the largest component of this network, consisting of 2.9~M nodes (which is 65\% of all nodes) and 12.5~M weighted links.

\item APS dataset comprises of over 450~k articles published in all journals of the American Physical Society since 1893 \cite{APS}. In this study, we use the preprocessed APS data \cite{2021ComPhysUbaldi} covering the period between January 1970 and December 2006 and containing 315~k documents with up to 11~co-authors, from which we built coauthorship network with the largest component consisting of 184~k nodes (which is 96\% of all APS authors recorded in this period) and 1~M weighted edges.  

\item CondMat is a weighted coauthorship network between scientists who published preprints on the Condensed Matter e-print archive between January 1995 and December 1999 \cite{2001PNASNewman}. To built the network we used a preprocessed bipartite dataset from \cite{CondMat}. The largest component of this network, which is taken into account for in-depth analysis, covers 14~k authors (which is 83\% of all authors) interacting via 45~k weighted links.
\end{itemize}

In this contribution, as in the previous one \cite{2022SciRepFronczak}, our main dataset is DBLP, from which we derive the key findings and to which we relate analysis made in the other two databases and in the synthetic model of scientific collaboration. The leading role of DBLP in our research is due not only to its largest size compared to the other two datasets. Rather, it results from the care of the authors of this database to disambiguate the names of the authors of publications \cite{2012Tang}. In DBLP, author names are disambiguated by the combination of algorithms and human curation \cite{2017Muller}, and not, as in many other bibliographic data - including APS and CondMat - represented by a string of characters corresponding to the surname(s) and initials of all forenames (or only the first one), which can lead to ambiguity of authors through merging or splitting their output \cite{2019JASISTKim}. However, the two additional datasets (APS and CondMat), although smaller and less accurate than DBLP, allow us to expand the scope of performed analyses by testing noise-resilience (e.g. due to incomplete data and problems with disambiguation of authors' names) of the recently observed weight-topology patterns \cite{2022SciRepFronczak}.

\subsubsection{Coauthorship network model}\label{xxx1}

From various different models of scientific collaboration proposed so far (see e.g. \cite{2002PhysABarabasi, 2005ScienceGuimera, 2013SciRepSun, 2018JASISTZhang}), for the analysis presented in this paper, we have chosen the model introduced in Ref.~\cite{2014PREKe}. The choice of this particular model was dictated by several reasons. First, the model reproduces the weight-topology correlations observed in real networks, which we wanted to address and comment on in this paper. Second, despite its simplicity, the model takes into account many important features of the evolution of real scientific collaboration networks that can be easily verified by examining readily available coauthorship networks. These features include: i. growth over time by adding new nodes - students, ii. emergence of new research groups, in which junior scientists (former students) become group leaders (the evolution of career stages \cite{2021JoILu}), iii. creation of new publications based on intra- and inter-group relations, and finally iv. high probability that the young scientist will give up a further scientific career. 

As indicated above, in the model studied, nodes are assigned to specific research groups in which they perform various functions. More specifically, each group consists of exactly one leader and a number of students, with the latter being "active" or "inactive" depending on the time elapsed since they were added to the network. It is assumed that the group leader is established at the time of group formation and remains in function until the end of the network evolution. The situation of students is a bit more complicated. After a node is added to the network, it is assigned to one of the existing research groups as its active student, who can participate in scientific research and coauthor publications. However, just like in the real world, after some time such a student may cease to be active, giving up further research activity, or may pursue a scientific career as a leader of a new research group.

In the considered model, inter-node connections result from common (i.e. coauthored) publications, the number of which translates into the edge weight. The model has two mechanisms of producing new publications, through intra- or inter-group collaboration, with the number of coauthors taken from a certain distribution $P(l)$. This distribution and any other parameters of the model are determined on the basis of real data. We comment on them later in the text, in the part devoted to simulation results.

To be more specific, the evolution of the network model under study proceeds as follows: 

(0) \textit{Beginning of the evolution:} The network starts to grow with a single research group consisting of a leader and one student. 

Then, in successive time steps, the following actions are performed (cf. description given in Ref. \cite{2014PREKe}):

(1) \textit{Intra-group publications:} With probability $c$, each group publishes one paper by itself. The paper is written by the group's leader and $l-1$ active students preferentially chosen from the same group based on student's scientific expertise, which corresponds to the length of student's activity period. 

(2) \textit{Inter-group publications:} Each group may publish up to $\alpha$ papers with another group. External collaboration always takes place with the same group, which is randomly chosen right after a new group appears. Same as for intra-group publications, each of these $\alpha$ papers is realized with probability $c$ and is coauthored by $l$ individuals (two leaders and $l-2$ students, which are preferentially chosen from the pool of all active students of the two groups).

(3) \textit{Groups' resource update:} Active students whose activity period has exceeded the threshold value $G$, with probability $f$ become leaders of new research groups, and with probability $1-f$ become inactive and no longer participate in network dynamics. A new active student is added to each group.

Although the model under study has several free parameters, the values of most of them can be approximated from various real data. For example, since in this paper we primarily use the coauthorship network extracted from DBLP Computer Science Bibliography to compare with the model results, we assume the distribution  $P(l)$ of the number of coauthors in publications consistent with the corresponding distribution in the mentioned database \footnote{By drawing the number $l$ of coauthors, we limit the range from which we randomly select to the size of the group (i.e. $G+1$ for intragroup publications and $2(G+1)$ for intergroup publications). For the value of $G=7$ adopted in this study, this limitation makes sense because, in the database under consideration, less than one per mille of publications has more than $2(G+1)=16$ authors.} (alternatively, one could use real data to train a model describing the distribution of coauthors in a similar way to \cite{2020JoIXie}), see Fig.~\ref{fig1}(a). Furthermore, although we examined a wide range of model parameter values in our studies, we ultimately decided to keep the values provided in Ref. \cite{2014PREKe} (where the model was originally introduced), as they result in the best agreement between simulations and real data. In Ref.~\cite{2014PREKe} the following values have been taken as a reference: $c=0.4$ for the probability to publish a paper; $\alpha=3$ for the number of inter-group publications; $f=0.2$ for the probability of a student becoming a group leader; and finally $G=7$ for the length of the students' activity period, which also determines the maximum number of active students in a research group. The rationale for these parameters is described in more detail in Ref.~\cite{2014PREKe}, to which we refer interested readers (for a broader perspective, see also the recent studies: \cite{2018JASISTZhang, 2018JASISTBu, 2015JASISTKim}).

\subsection{Results}

\subsubsection{Real data vs. simulation results}

\begin{figure*}[t]
	\includegraphics[width=2\columnwidth]{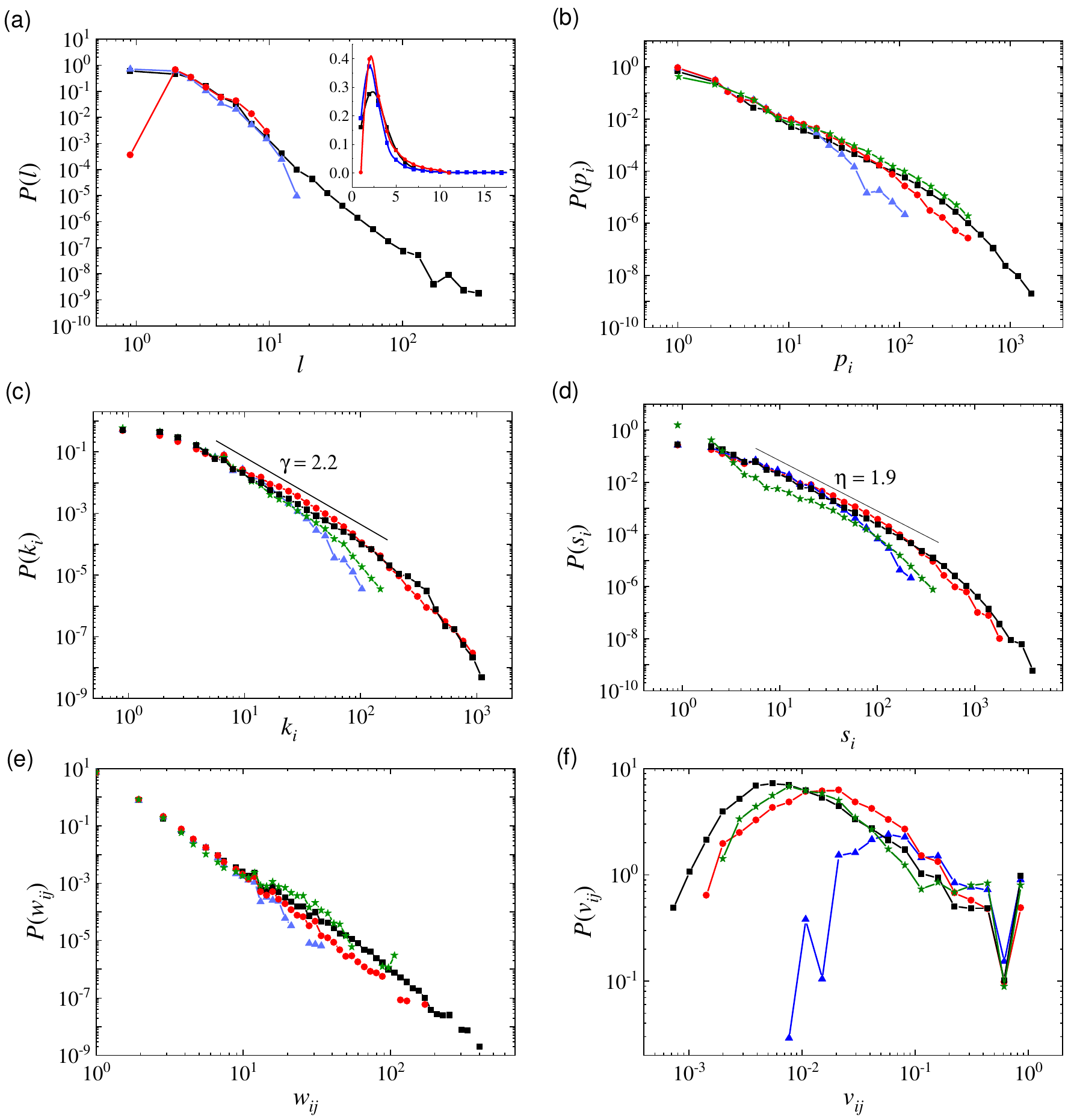}
	\caption{\textbf{Structural properties of coauthorship networks.} The following graphs show distributions of: (a) the number of coauthors per paper, $P(l)$; (b) the number of publications per author, $P(p_i)$; (c) the node degrees, $P(k_i)$; (d) the node strengths, $P(s_i)$; (e) the edge weights (or symmetric tie strengths), $P(w_{ij})$; (f) the asymmetric tie strengths, $P(v_{ij})$. The symbols used are: black squares for DBLP, red circles for APS, blue triangles for CondMat, and green stars for results of numerical simulations obtained from the model network. \label{fig1}}
\end{figure*}

In Fig.~\ref{fig1}(b)-(f), we show basic structural characteristics of real coauthorship networks (DBLP, APS, and CondMat) and the model network with $N\simeq 10^4$ nodes (averaged over $100$ realizations). Two obvious conclusions arise after analysing this figure. First: The examined features of real networks are very similar to each other. It is reasonable to claim that the slight differences in the range of the data shown are primarily related to the size of the analysed networks, which varies from millions of nodes (in DBLP), through hundreds (in APS) to tens of thousands (in CondMat). Second: The synthetic model of scientific collaboration reflects very well the basic features of the reference coauthorship networks, including their skewed distributions of node degrees, $P(k_i)$, and strengths, $P(s_i)$ (where the node strength is given as the sum of the weights of its edges: $s_i=\sum_jw_{ij}$), as well as the fat-tailed distributions of edge weights (tie strengths) $P(w_{ij})$ and $P(v_{ij})$ (where $w_{ij}$ represents the number of joint papers, and $v_{ij}=w_{ij}/p_i\neq v_{ji}$ (\ref{vij}) stands for an asymmetric tie strength, which is discussed afterwards). The good agreement between the model and real data, as can be seen in this figure, is all the more convincing as we checked that the differences between them decreased as the size of the model networks increased. The above results make the considered model a promising test-bed to study weight-topology correlations in scientific collaboration networks, which enables the formulation of well-established conclusions, based not only on real datasets but also on results of repeatable numerical simulations.

\subsubsection{Granovetter's hypothesis in coauthorship networks}

\begin{figure*}[t]
	\includegraphics[width=2\columnwidth]{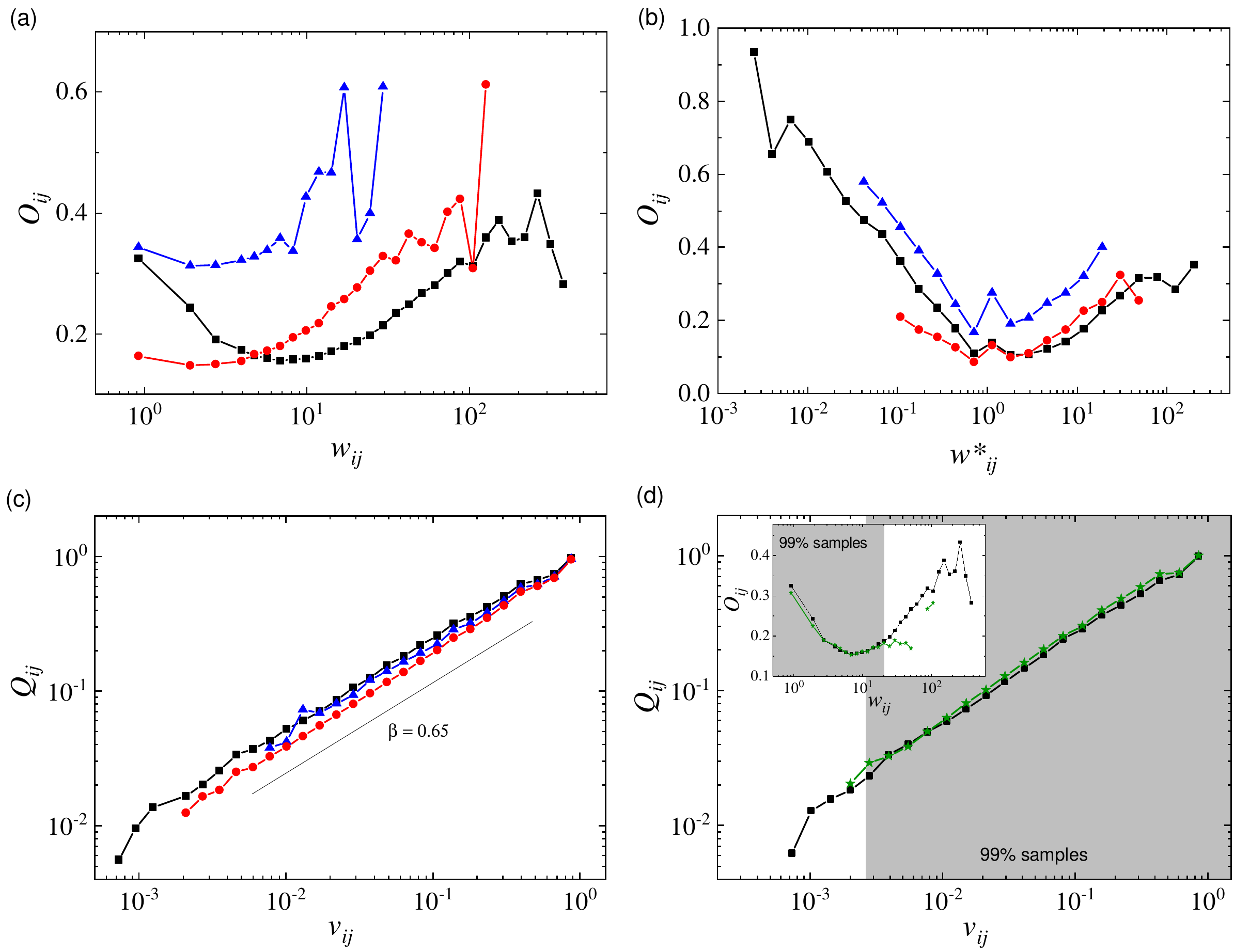}
	\caption{\textbf{Weight-topology correlations in coauthorship networks} as observed on the basis of relationship between variously defined tie strengths ($w_{ij}$, $w^*_{ij}$, and $v_{ij}$) and neighbourhood overlaps ($O_{ij}$ and $Q_{ij}$). Detailed description of the notation used is given in the main test. Graphical symbols used are the same as in Fig.~\ref{fig1}, i.e. black squares for BDLP, red circles of APS, etc. Note that the panels (a), (b), and (c) show only real coauthorship networks. Panel (d) presents results of numerical simulations of the synthetic network model and DBLP data for comparison.\label{fig2}}
\end{figure*}

As mentioned in the introduction, according to the Granovetter's hypothesis, strong social ties are expected to be associated with densely connected groups of individuals, while weaker ties act as bridges between these groups. To quantitatively characterize such weight-topology correlations, in Ref.~\cite{2007PNASOnnela}, the relationship between tie strength, $w_{ij}$, connecting two nodes ($i$ and $j$) and their neighbourhoods' overlap, $O_{ij}$, has been used, with the overlap defined as the ratio of the number of common neighbours, $n_{ij}$, of this node pair to the number of all their neighbours:
\begin{equation}\label{Oij}
	O_{ij}=\frac{n_{ij}}{(k_i-1)+(k_j-1)-n_{ij}}.
\end{equation}
Correspondingly, clear empirical support for the Grannovetter's hypothesis, manifested a monotonically increasing dependence between $w_{ij}$ and $O_{ij}$, has indeed been observed in many social networks, but not in coauthorship networks (see Fig.~\ref{fig2}(a) and (d - inset graph)), making the latter a flagship example of systems in which the hypothesis fails. 

Interestingly, the characteristic U-shape non-monotonic relation between tie strength and symmetric overlap $O_{ij}$, which is interpreted as the evidence of atypical weight-topology correlations in scientific collaboration networks becomes even more apparent, when the Newman's definition \cite{2001PREbNewman, 2012EPLPan, 2014PREKe} of tie strength, $w_{ij}^*$, is taken into account (see Fig.~\ref{fig2}(a),(b)). To grasp the difference between $w_{ij}$ and $w^*_{ij}$, recall that the tie strength $w_{ij}$, which is the standard used throughout this paper, stands for the number of joint publications, that is also the number of times a collaboration between two scientists has been repeated. Correspondingly, the Newman's tie strength is defined as: 
\begin{equation}
w^*_{ij}=\sum_{p_{ij}}\frac{1}{l_{ij}-1}\neq \sum_{p_{ij}}1=w_{ij},
\end{equation} 
where the sum runs over the set of papers $p_{ij}$ co-authored by $l_{ij}$ scientists, including $i$ and $j$. The motivation behind the Newman's formula for $w^*_{ij}$ is that an author divides his/her time and other resources between $l_{ij}-1$ collaborators, and thus the tie strength of such a collaboration should vary inversely with $l_{ij}-1$.  

The possible cause of the failure of the Granovetter's hypothesis in scientific collaboration networks has only recently been clarified in Ref.~\cite{2022SciRepFronczak}, where it was suggested that the non-monotonic $O_{ij}(w_{ij})$ and/or $O_{ij}(w^{*}_{ij})$ relations characterizing these networks are due to the definition of the neighbourhood overlap, Eq.~(\ref{Oij}) (hereafter called \textit{symmetric overlap}) which is not properly suited to be a local network measure in networks with scale-free node degree-distributions. 

The above mentioned problem with the symmetric overlap is particularly acute in the case of links connecting nodes with significantly different degrees. In such cases, for $k_i\ll k_j$, Eq.~(\ref{Oij}) can be simplified to $O_{ij}\simeq n_{ij}/k_j$, which shows that it is strongly biased towards nodes with high degrees, distorting the image of the common neighbourhood as seen from the perspective of nodes with small degrees. This drawback of symmetric overlap gains importance in networks with highly skewed, fat-tailed node degree distributions $P(k_i)$. In such networks, as brilliantly exploited by the degree-based mean-field theory of complex networks \cite{HMF1, HMF2, HMF3}, node degree distributions for nearest neighbours are even more fat-tailed than the original distributions $P(k_i)$. As a result, the number of edges in such networks connecting nodes with high and low degrees can be very high, leading to an unintended overrepresentation of strongly connected nodes by Eq.~(\ref{Oij}). 
 
\begin{figure*}[t]
	\includegraphics[width=1.5\columnwidth]{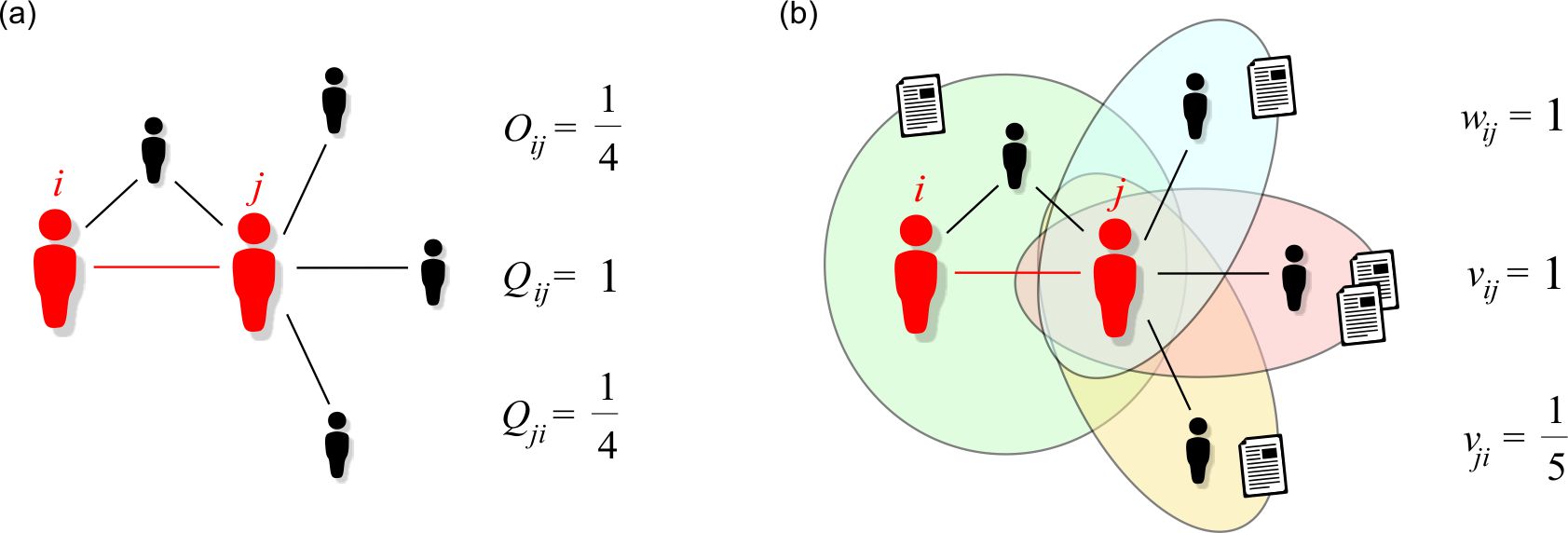}
	\caption{\textbf{a) Difference between symmetric and asymmetric neighbourhood overlap}. In the figure, a pair of scientists ($i$ and $j$) is shown, with different numbers of collaborators (respectively, $k_i=2$ and $k_j=5$) and one neighbour in common (i.e. $n_{ij}=1$). The figure shows that even in such a simple situation the neighbourhood overlap, as seen from the perspective of each of the two nodes, is significantly different: $Q_{ij}<Q_{ji}=O_{ij}$. \textbf{b) Difference between symmetric and asymmetric tie strength}. The figure presented in part a) is supplemented with additional data, which allow to determine appropriate edge weights. In particular, it can be seen from this figure that the node $i$ stands for an individual who coauthored (with two other individuals, including $j$) only one paper. The corresponding tie strengths are: $v_{ij}=w_{ij}>v_{ji}$. \label{fig3}}
\end{figure*}

To overcome the aforementioned problems with the symmetric overlap $O_{ij}$, the concept of \textit{asymmetric overlap} has been introduced in Ref.~\cite{2022SciRepFronczak}:
\begin{equation}\label{Qij}
	Q_{ij}=\frac{n_{ij}}{k_i-1}\neq Q_{ji}, 
\end{equation}
and it was used to describe the overlap between the neighbourhoods of two connected nodes from the perspective of each node separately \footnote{Note that the definition of the asymmetric neighbourhood overlap, Eq.~(\ref{Qij}), that we use in our manuscript is similar to the so-called \textit{edge clustering coefficient}: $C_{ij}=n_{ij}/\mbox{min}[k_i-1,k_j-1]$ \cite{2004PNASRadicchi}. However, the difference between the two measures is essential, because $Q_{ij}\neq Q_{ji}$ while $C_{ij}=C_{ji}$.}. In the context of coauthorship  networks, this new definition is free from the shortcomings of the previous one. In particular, it copes well with collaborating scientists whose degrees (ego-networks) differ significantly in size - that is, when their common neighbours (if any) are a significant part of the neighbourhood of one node and an insignificant part of the neighbourhood of the other. The relevant situation is illustrated in Fig.~\ref{fig3}(a).

The concept of asymmetric overlap naturally leads to the idea of directed networks and justifies the introduction of the \textit{asymmetric tie strength}:
\begin{equation}\label{vij}
	v_{ij}=\frac{w_{ij}}{p_{i}}\neq v_{ji},
\end{equation} 
where $p_i$ stands for the number of all publications of the $i$-th author (note that the number of publications does not have to be equal to the strength of the node: $p_i\neq s_i=\sum_j w_{ij}$). The intuitive rationale behind Eq.~(\ref{vij}) is illustrated in Fig.~\ref{fig3}(b) and it proceeds as follows: For a young scientist, with a small number of publications, each publication makes a significant contribution to his or her publication output, just as each coauthor is an important part of his or her research environment (cf. Eqs.~(\ref{Qij}) and~(\ref{vij})). However, the importance of each publication and collaboration from the perspective of an established scientist with a large number of publications and an extensive network of collaborators is completely different. Depending on the circumstances, a given number of joint publications (e.g., $w_{ij}=1$) may have a completely different meaning. 

In Fig.~\ref{fig2}(c) and (d-main panel), relationship between asymmetric neighbourhood overlap, $Q_{ij}$, is shown against the corresponding asymmetric tie strength, $v_{ij}$. Remarkably, although the relationships with the use of symmetric network measures (see Fig.~\ref{fig2}(a),(b) and (d - inset graph)) are cumbersome to interpret (e.g. due to U-shape relation observed for $O_{ij}(w^*_{ij})$ and significant differences between the real data observed for  $O_{ij}(w_{ij})$), the relationship between asymmetric measures $v_{ij}$ and $Q_{ij}$ seems to be universal for all studied networks. The reasonable explanation for this observation is that, the relationship between symmetric tie strengths, $w_{ij}$ and $w^*_{ij}$, and the symmetric overlap $O_{ij}$ is not an informative measure for weight-topology correlations in coauthorship networks. On the other hand, the result for asymmetric measures is of particular importance as it aspires to be a universal scaling law that would require verification in other social networks, not only collaborative. 

In fact, Fig.~\ref{fig2}(c) confirms that the Granovetter's hypothesis holds in coauthorship networks. In other words, from the point of view of an individual scientist, strong ties do really correspond to dense local neighbourhoods, contrary to what has been suggested in other studies on these networks. The perspective of a single node, being one of the two ends of an edge, is important here. The new measures introduced ($Q_{ij}$ and $v_{ij}$) quantitatively capture the so far elusive concept of relativity in social relations. Following this line of reasoning, it may be tempting to say that the measured absolute tie strength (i.e. $w_{ij}$) is a kind of compromise and depends on relative strengths of the tie as seen from its both ends (i.e. $v_{ij}$ and $v_{ji}$). Moreover, it seems reasonable that similar thinking should also apply to the connection probability, not just to its weight. In this context, it is surprising that none of the so far proposed network measures that are used in link-prediction methods take into account, at least not explicitly, the asymmetry of these links. In the rest of the publication, we refer to these issues.

\section{Asymmetry-based link prediction}\label{part2}

\subsection{Methods}

Link prediction refers to the problem of finding missing or hidden links that are likely to exist in networks or will appear there in the near future \cite{2011PhysALu, 2017ACMMartinez}. Predicting new friendships in social media or new collaborations in coauthorship networks \cite{2007Kleinberg, 2007Schafer, 2014SciGuns}, discovering previously unknown interactions in biological networks \cite{2008NatureClauset}, predicting scientific research trends \cite{2020JoIBehrouzi}, or providing bibliography recommendations \cite{2022SciPornprasitSci} are a few examples showing the importance and the diversity of the applications that can benefit from link prediction. It is also worth noting that link prediction is not limited to single-layer networks and prediction methods can utilise data from multiple layers \cite{2021JoIKarimi} representing various types of interactions.

The simplest predicting methods are based on nodes' neighbourhood-related structural information that is used to compute the so-called similarity score, $s_{ij}$, of each pair of nodes in the network. Then, by ranking the pairs based on this score, an inference is made as to the existence or absence of edges. In the literature on link prediction, one can find dozens of such scores (or indicators) that perform better or worse depending on the network under study. In particular, the following examples of symmetric indicators have been widely employed by the research community due to their simplicity, computational efficiency and performance: the common neighbours index \cite{2001PRENewman}, the Jaccard’s index \cite{1901Jaccard}, the Adamic-Adar (AA) index \cite{2003SocNetwAdamic}, and the resource allocation (RA) index \cite{2009EPJBZhou}. Further in this subsection, using the Jaccard's index as an example, we will show that by redefining this index to take into account the asymmetry of network connections, one can significantly increase its prediction efficiency. The perspective will also allow us to comment on the surprisingly high performance of the RA index, pointing to the asymmetry as a promising direction for further research on effective prediction methods.

The Jaccard's index is widely used in information retrieval systems to compare the similarity and diversity of sample sets. In the context of link prediction methods, the index measures the proportion of common neighbours of two nodes ($i$ and $j$) in the total number of their neighbours. Correspondingly, it is given by the expression:
\begin{equation}\label{Jaccard}
	\mbox{JC}_{\,ij}=\frac{|\Gamma(i)\cap\Gamma(j)|}{|\Gamma(i)\cup\Gamma(j)|},
\end{equation}
where $\Gamma(i)$ is the set of nearest neighbours of $i$, and $|\Gamma(i)|$ is the cardinality of this set. Since $|\Gamma(i)|=k_i$ and $|\Gamma(i)\cap\Gamma(j)|=n_{ij}$, the above formula can be rewritten as
\begin{equation}\label{Jaccard1}
	\mbox{JC}_{\,ij}=\frac{n_{ij}}{k_i+k_j-n_{ij}}\simeq O_{\,ij},
\end{equation}
which shows that the definition of the Jaccard's index is almost identical to the definition of the symmetric neighbourhood overlap, cf. Eq.~(\ref{Oij}). 

Having the similarity score $s_{ij}$ defined (e.g. $s_{\,ij}=\mbox{JC}_{\,ij}$), the link prediction proceeds as follows: One ranks the non-connected pairs of nodes in descending order according to the values of $s_{ij}$ and then the pairs for which the score exceed some established threshold $T$ are considered to be connected. At this point, at least two problems arise, that we comment on below.

First, to validate the ranking method used and to evaluate the similarity score chosen, one has to know \textit{a~priori} which identified links are indeed present in the network. Thus, for the testing purposes, one has to construct a~set of node pairs which include those pairs that are connected in the original network (labelled as positive links -- P) and those that are not (labelled as negative links -- N). Construction of such a set is not a trivial task per se, because the studied dataset is highly imbalanced - the number of negative (i.e non-existing) links is much larger than the number of positive ones. Moreover most of node pairs have no common neighbours, which may result in their similarity scores equal to zero (e.g. $\mbox{JC}_{\,ij}=0$ for $n_{\,ij}=0$). To overcome this problem, in this study, we construct our testing set by selecting $d$ existing links and $d$ non-existing links, both from those node-pairs which share at least one common neighbour. 

\begin{figure*}
	\includegraphics[width=1.8\columnwidth]{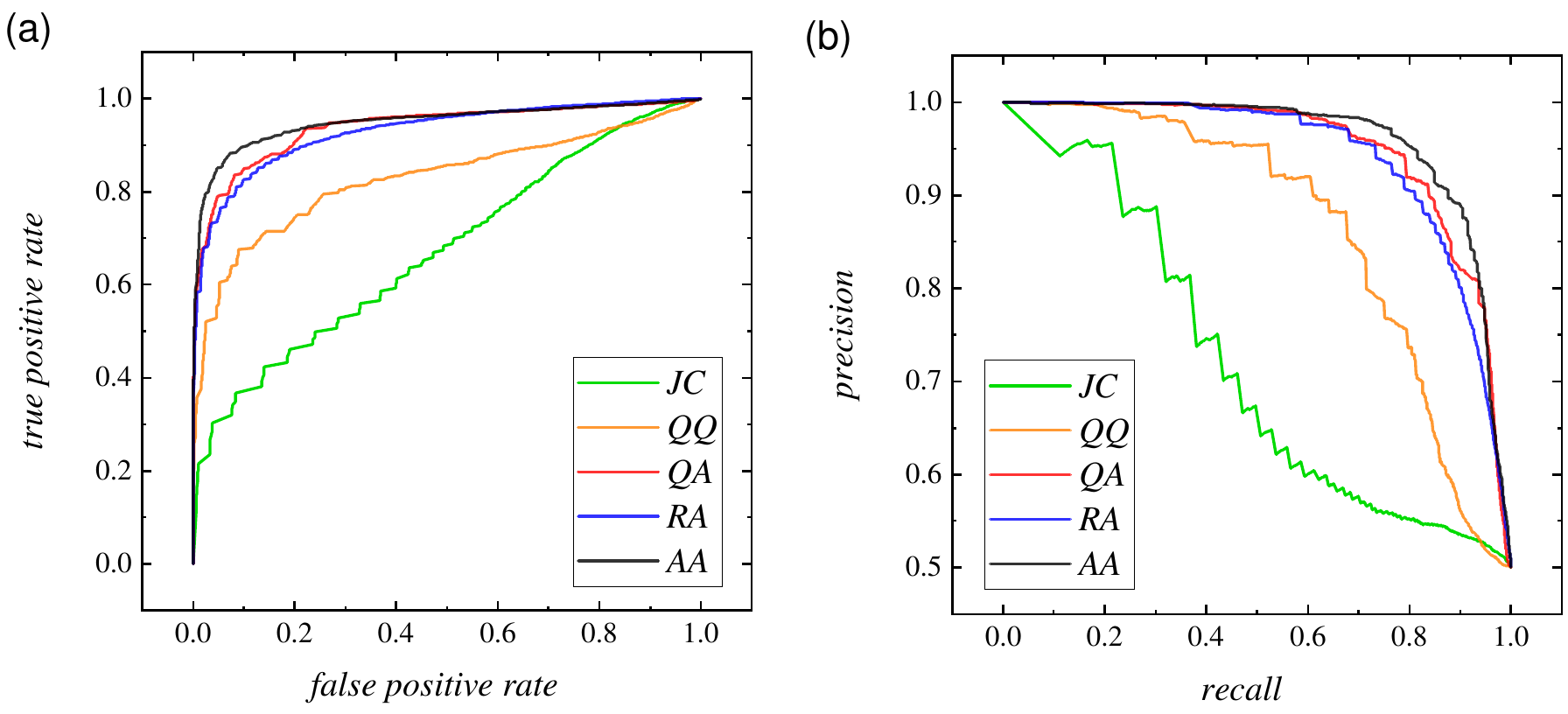}
	\caption{\textbf{Performance of similarity-based link prediction methods in coauthorship networks}. ROC curves (a) and precision-recall curves (b) for different score measures in the DBLP dataset.\label{fig4}}
\end{figure*}

Given such a correctly balanced testing set of existing and non-existing links, one can perform ranking on this set and create the confusion matrix (see Tab. I), whose elements - representing the numbers of connections labelled as: true positive (TP), true negative (TN), false positive (FP), and false negative (FN) - allow to derive several useful metrics to asses performance of a link-prediction method. In what follows, we use three, perhaps the most commonly used, metrics (see axis description in Fig.~\ref{fig4}): 

(1) \textit{sensitivity}, also referred to as recall or the true positive rate (TPR), which measures the proportion of existing links that are correctly identified as true positive matches to all existing links: TPR=TP/(TP+FN),

(2) \textit{specificity}, also known as the true negative rate (TNR), measures the proportion of non-links that are correctly identified as negative links: TNR=TN/(TN+FP), and finally

(3) \textit{precision}, also called the link accuracy, that is the proportion of all correctly identified links to all node-pairs in the set that are classified (correctly or not) as existing links: PR=TP/(TP+FP). 

\begin{table}[]\label{tab1}
	\begin{tabular}{l|c|c|}
		\cline{2-3}
		& predicted links & predicted non-links  \\ \hline
		\multicolumn{1}{|l|}{positive links}& True Positives ($\mbox{TP}$) & False Negatives ($\mbox{FN}$)  \\ \hline
		\multicolumn{1}{|l|}{negative links}& False Positives ($\mbox{FP}$) & True Negatives ($\mbox{TN}$)  \\ \hline
	\end{tabular}\caption{Confusion matrix for link prediction.}
\end{table}

Although the above measures seem similar, they relate to different aspects of link prediction and thus are complementary to each other. In particular, precision evaluates the correctness, while recall evaluates the completeness of both the similarity score and the method used. Generally, there is a trade-off between precision and recall, whereby a larger threshold $T$ increases precision and decreases recall.

The second problem with the method described is that the above measures use a fixed threshold to rank node pairs, but the value of this threshold may not be necessarily available or be the most optimal one. For example, it may be domain dependent. To deal with such cases threshold curve based metrics are used. This is where one of the most important goals of link prediction research emerges, which is optimizing these curves to find the most effective similarity score, $s_{ij}$. It is also where our research on the role of asymmetry in social ties contributes to the already huge research area of link prediction.

There are generally two such curves in use. The first one, the ROC curve (from: receiver operating characteristic) represents the performance trade-off between true positives TPR and false positives FPR=$1-$TNR at different decision boundary thresholds. It can be interpreted as the probability that a randomly chosen true positive link will be ranked higher than a randomly chosen true negative link \cite{1982Hanley}. The area under the ROC curve, AUC, is always between 0 and 1, and, generally, the performance of any realistic classifier at AUC measure should be higher than 0.5 (which corresponds to completely random classifier) \cite{2002Mason}. The second curve, the PR curve (from: precision recall) considers only prediction of the positive links, which, in some situations, can be more useful and informative, e.g. when negative links are not interesting \cite{2006Davis}. Similarly to the AUC value characterizing the ROC curve, to get one number that describes performance of the method, one can also calculate the PRAUC, being the area under the precision-recall curve. Such a single value can be understood as the average of precision scores calculated for each recall threshold. The higher values of both threshold curve metrics, AUC and PRAUC, correspond to those similarity scores, $s_{ij}$, that rank better the pairs of nodes towards the discovery of existing edges between them.

\subsection{Results}

\subsubsection{Similarity scores accounting for link asymmetry}

\begin{table*}[]\caption{Performance results of analysed score measures tested on the DBLP, APS, CondMat datasets and the discussed model of scientific collaboration network.}
		\begin{tabular}{V{3}cV{3}ccV{3}ccV{3}ccV{3}ccV{3}}
			\clineB{1-9}{3}
			\multicolumn{1}{V{3}cV{3}}{\multirow{2}{*}{similarity score}}  &\multicolumn{2}{cV{3}}{DBLP} &\multicolumn{2}{cV{3}}{APS} &\multicolumn{2}{cV{3}}{Cond-Mat}     &\multicolumn{2}{cV{3}}{model}  
			\\ \cline{2-9}
			\multicolumn{1}{V{3}cV{3}}{} &\multicolumn{1}{c|}{AUC}  & PRAUC &\multicolumn{1}{c|}{AUC}  & PRAUC &\multicolumn{1}{c|}{AUC}  & PRAUC &\multicolumn{1}{c|}{AUC}  & PRAUC \\ \clineB{1-9}{3}
			\begin{tabular}[c]{@{}l@{}} JC \end{tabular}   &\multicolumn{1}{c|}{0.684} & 0.714  &\multicolumn{1}{c|}{0.698} & 0.684  &\multicolumn{1}{c|}{0.764} & 0.766  &\multicolumn{1}{c|}{0.634} & 0.659  \\ \hline
			\begin{tabular}[c]{@{}l@{}} QQ \end{tabular}  &\multicolumn{1}{c|}{0.829} & 0.865  &\multicolumn{1}{c|}{0.758} & 0.795  &\multicolumn{1}{c|}{0.847} & 0.870  &\multicolumn{1}{c|}{0.798} & 0.823  \\ \hline
			\begin{tabular}[c]{@{}l@{}} QA \end{tabular}&\multicolumn{1}{c|}{0.939} & 0.949  &\multicolumn{1}{c|}{0.925} & 0.941  &\multicolumn{1}{c|}{0.901} & 0.920  &\multicolumn{1}{c|}{0.860} & 0.885  \\ \hline
			\begin{tabular}[c]{@{}l@{}} RA \end{tabular}&\multicolumn{1}{c|}{0.934} & 0.945  &\multicolumn{1}{c|}{0.922} & 0.927  &\multicolumn{1}{c|}{0.910} & 0.924  &\multicolumn{1}{c|}{0.869} & 0.902  \\ \hline
			\begin{tabular}[c]{@{}l@{}} AA \end{tabular}&\multicolumn{1}{c|}{0.948} & 0.952  &\multicolumn{1}{c|}{0.937} & 0.944  &\multicolumn{1}{c|}{0.918} & 0.929  &\multicolumn{1}{c|}{0.888} & 0.920  \\ \clineB{1-9}{3}
	\end{tabular}
\end{table*}

The Jaccard's index, Eq.~(\ref{Jaccard}), which is equivalent to the symmetric overlap, Eq.~(\ref{Oij}), was one of the first similarity measures used in link prediction methods. Nevertheless, it was quickly realized that the link prediction based on this measure is not much better than a random classifier. Correspondingly, in the case of coauthorship networks analysed in this study, the values of AUC and PRAUC are close to 0.7, which is a rather poor result compared to other similarity scores reported in Tables~II and~III, which have values up to 0.97. Reasons of such a poor performance of the Jaccard's index, however, have not been thoroughly investigated. In this context, the results on weight-topology correlations in social networks presented in this paper shed some light on the problem. The relevant reasoning is as follows: If the tie strength of the missing link is high, the probability of its existence, based on the corresponding similarity score, should also be high. Likewise, low-strength ties should be less likely to be realized. Consequently, the prediction methods based on the Jaccard's index will have poor performance in networks where the correlation between tie strengths, $w_{ij}$, and the symmetric neighbourhood overlap, $O_{ij}$, are weakly positive or absent at all. This is the situation we deal with in social networks with fat-tailed degree distributions, of which coauthorship networks are a particularly vivid example. 

To support the above reasoning, we have introduced and analysed a simple similarity score:
\begin{equation}\label{QQ1}
	\mbox{QQ}_{\,ij}=\frac{|\Gamma(i)\cap\Gamma(j)|}{|\Gamma(i)|}+\frac{|\Gamma(i)\cap\Gamma(j)|}{|\Gamma(j)|},
\end{equation}
the form of which refers to the sum of the asymmetric neighbourhood overlaps (\ref{Qij}) of the considered pair of nodes: 
\begin{equation}\label{QQ}
	\mbox{QQ}_{\,ij}=\frac{n_{ij}}{k_i}+\frac{n_{ij}}{k_j}\simeq Q_{\,ij}+Q_{\,ji}.
\end{equation}
As one can see in Figs.~\ref{fig4} and~\ref{fig6} and Tab.~II, for all studied coauthorship networks (real and synthetic), the link prediction results obtained with this similarity score are much better than with the Jaccard's index, Eq.~(\ref{Jaccard1}). Moreover, the results can be further improved by making adjustments learned from the Adamic-Adar index (see Eq.~(\ref{AA})), in which the importance of nodes is expressed not by their degree, but by the logarithm of the degree: 
\begin{equation}\label{QA}
	\mbox{QA}_{\,ij}=\frac{n_{ij}}{\log k_i}+\frac{n_{ij}}{\log k_j}.
\end{equation}

As compared to other similarity scores listed in Tab.~II, the results obtained using the QA index are significant for at least two reasons. Firstly, the accuracy of predictions according to QA is similar to the accuracy obtained using measures such as the resource allocation index \cite{2011PhysALu}:
\begin{equation}\label{RA}
	\mbox{RA}_{\,ij}=\sum_{z\in\Gamma(i)\cap\Gamma(j)}\frac{1}{k_z},
\end{equation}
and the Adamic-Adar index:
\begin{equation}\label{AA}
	\mbox{AA}_{\,ij}=\sum_{z\in\Gamma(i)\cap\Gamma(j)}\frac{1}{\log k_z},
\end{equation}
both of which require much more detailed information about the neighborhood of the possible edge between $i$ and $j$ than just the number of their common neighbors, $n_{ij}=|\Gamma(i)\cap\Gamma(j)|$, as QA does. Secondly, it shows that taking the asymmetry of social ties into account can significantly improve the effectiveness of link prediction methods. In the rest of the paper, we refer to the last remark, showing that the high efficiency of RA and AA indicators in social networks is obviously related to the asymmetry of social ties. In showing this, in order to avoid distraction, we limit ourselves to the RA index only.

\subsubsection{Asymmetry-based perspective on RA-like scores}

All the local similarity indices we have discussed so far have one thing in common: They are designed based on the assumption that two nodes are more likely to have a link if they have many common neighbors. In particular, the RA index~(\ref{RA}) is the greater the more common neighbors the nodes $i$ and $j$ have, but it also reduces the contribution of high degree common neighbors by assigning more importance to those less-connected~\footnote{Note that the AA index also punishes the high-degree common neighbors but to a lesser extent than RA. It depends on the network under study which of the approaches to punish nodes with high degrees is better (RA or AA). In scientific collaboration networks we study in this paper, AA performs better, but this is not a rule, because in other networks (not only the social ones) it may be the other way around \cite{2011PhysALu,2016SciRepZhu}.}. In what follows, we show that the punishment of high degree nodes in Eq.~(\ref{RA}) can be justified by the strong triadic closure property, assuming that the property takes into account the asymmetry of social ties.

\begin{figure}
	\includegraphics[width=0.9\columnwidth]{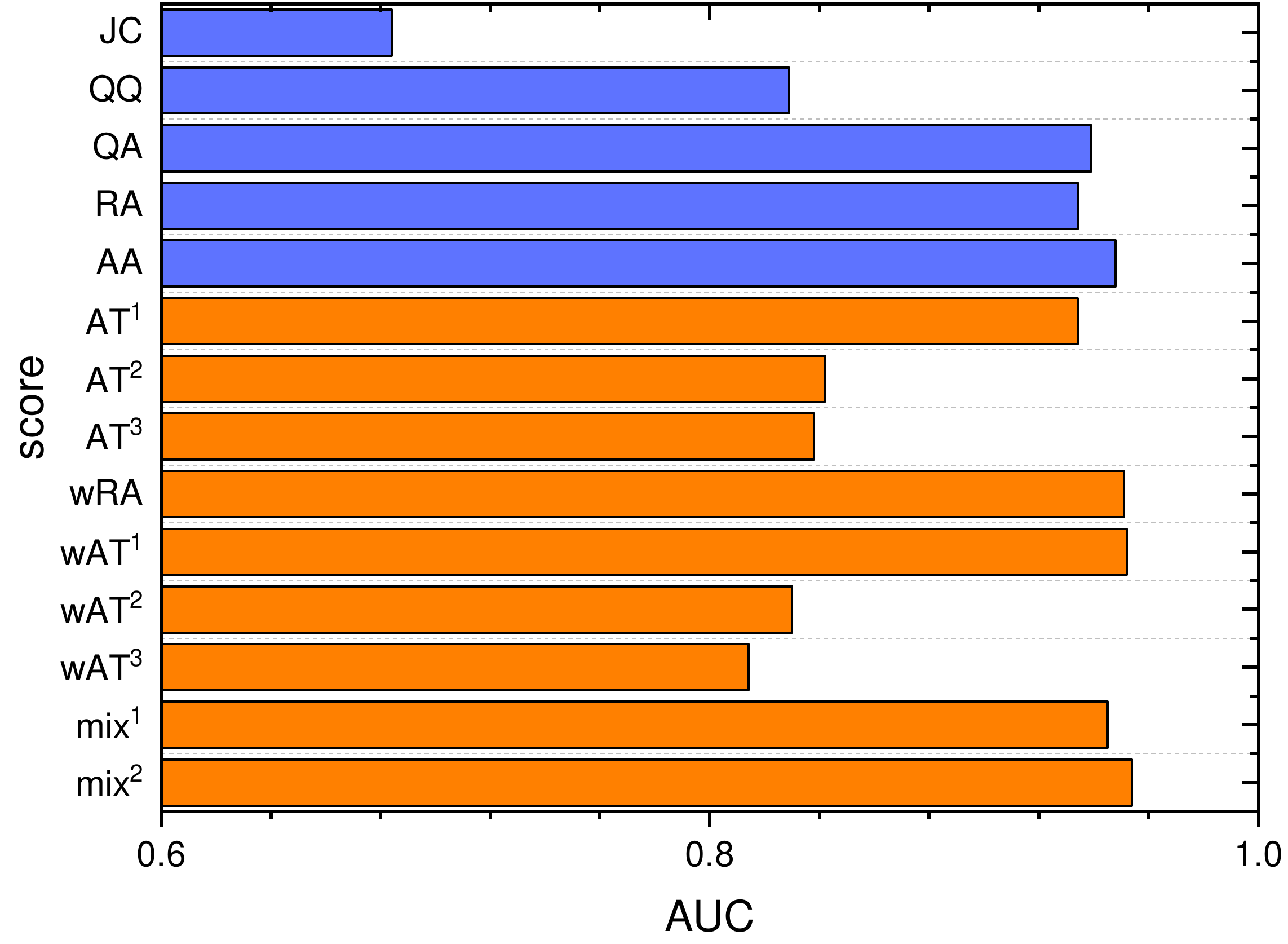}
	\caption{\textbf{AUC measures for different scores} calculated for DBLP dataset. Blue and orange colors correspond to the data presented in the Table II and III respectively.}\label{fig6}
\end{figure}

Triadic closure property states that: If two people $i$ and $j$ in a social network have a friend $z$ in common, then there is an increased \textit{likelihood }that they will become friends themselves at some point in the future (see \cite{2010bookEasley}, p.~44). Accordingly, the strong triadic closure property completes the previous statement saying that: If a node $z$ has edges to $i$ and $j$, then the connection between $i$ and $j$ is \textit{especially likely} to form if $z$'s edges to $i$ and $j$ are both \textit{strong ties} (see \cite{2010bookEasley}, p.~49).

Clearly, both triadic closure properties are intuitively very natural. Furthermore, although their original wordings apply to single triads, i.e. those in which nodes $i$ and $j$ have only one common neighbor $z$, it is reasonable to assume that the more common neighbors, the higher the \textit{likelihoods} in question should be. In the context of link predictability - the primary concern of which is to correctly estimate the mentioned likelihoods - the above reasoning leads to the simplest similarity measures defined as a bare number of common neighbors \cite{2011PhysALu, 2016SciRepZhu}: 
\begin{equation}\label{CN}
	\mbox{CN}_{\,ij}=n_{\,ij},
\end{equation}
and its weighted version:
\begin{equation}\label{wCN}
	\mbox{wCN}_{\,ij}=\sum_{z\in\Gamma(i)\cap\Gamma(j)}(w_{zi}+w_{zj}),
\end{equation}
where $w_{ij}$ stand for the symmetric tie strength between $i$ and $j$. However, as numerous studies show, these indices are not as efficient as the previously introduced RA (\ref{RA}) and AA (\ref{AA}) scores. Below we show, where this inefficiency comes from and how to improve it without going beyond the concept of triadic closure.

\begin{figure}
	\includegraphics[width=0.8\columnwidth]{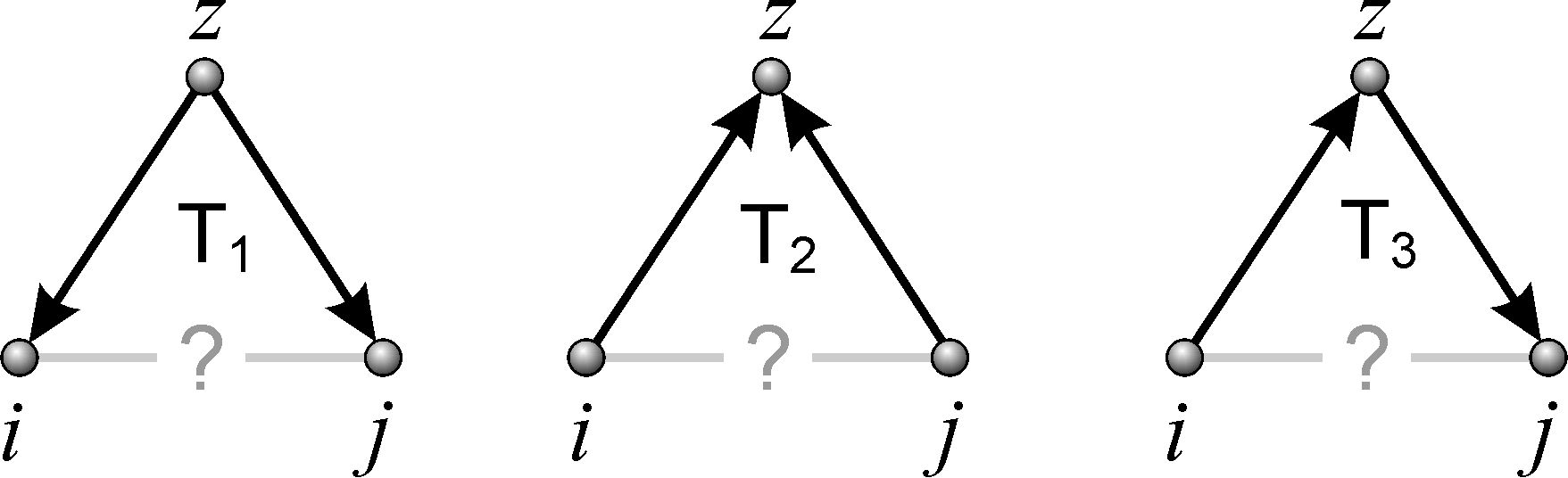}
	\caption{\textbf{Three types of directed triads} that are analysed in the text.}\label{fig5}
\end{figure}

The idea - which is consistent with our findings about asymmetry of social ties - is to replace the symmetric weights $w_{ij}$ in Eq.~(\ref{wCN}) with the asymmetric tie strengths $v_{ij}$~(\ref{vij}) or with the asymmetric neighbourhood overlaps $Q_{ij}$~(\ref{Qij}) that show a high correlation with asymmetric tie strengths. In the case of binary networks, the corresponding similarity score - referring to the newly addressed concept of the directed triad closure \cite{2019ProcYin,2020NetSciYin} - can be defined as:     
\begin{eqnarray}\label{AT1}
	\mbox{AT}^1_{\,ij}&=&\sum_{z\in\Gamma(i)\cap\Gamma(j)}(Q_{zi}+Q_{zj})\\ \label{AT1a} &=&\sum_{z\in\Gamma(i)\cap\Gamma(j)}\frac{n_{zi}+n_{zj}}{k_z-1},
\end{eqnarray}
and for the weighted networks, as:
\begin{eqnarray}\label{wAT1}
	\mbox{wAT}^1_{\,ij}&=&\sum_{z\in\Gamma(i)\cap\Gamma(j)}(v_{zi}+v_{zj}),
\end{eqnarray}
where the acronyms AT and wAT account for 'asymmetric triad' and 'weighted asymmetric triad', and the superscript - here '1' - refers to the triad ordering T$_{1}$ as shown in  Fig.~\ref{fig5}.

\begin{table*}[]\caption{Performance results of analysed RA-like similarity measures tested on the DBLP, APS, CondMat datasets and the discussed model of scientific collaboration network.}
		\begin{tabular}{V{3}cV{3}ccV{3}ccV{3}ccV{3}ccV{3}}
			\clineB{1-9}{3}
			\multicolumn{1}{V{3}cV{3}}{\multirow{2}{*}{similarity score}}  &\multicolumn{2}{cV{3}}{DBLP}         &\multicolumn{2}{cV{3}}{APS}          &\multicolumn{2}{cV{3}}{Cond-Mat}     &\multicolumn{2}{cV{3}}{model}   \\ \cline{2-9}
			\multicolumn{1}{V{3}cV{3}}{}                                   &\multicolumn{1}{c|}{AUC}  & PRAUC &\multicolumn{1}{c|}{AUC}  & PRAUC &\multicolumn{1}{c|}{AUC}  & PRAUC &\multicolumn{1}{c|}{AUC}  & PRAUC \\ \clineB{1-9}{3}
			\begin{tabular}[c]{@{}l@{}} RA \end{tabular}&\multicolumn{1}{c|}{0.934} & 0.945  &\multicolumn{1}{c|}{0.922} & 0.927  &\multicolumn{1}{c|}{0.910} & 0.924  &\multicolumn{1}{c|}{0.869} & 0.902  \\ \hline
			\begin{tabular}[c]{@{}l@{}} AT$^{1}$ \end{tabular}   &\multicolumn{1}{c|}{0.934} & 0.949  &\multicolumn{1}{c|}{0.924} & 0.941 &\multicolumn{1}{c|}{0.883} & 0.915  &\multicolumn{1}{c|}{0.874} & 0.911 \\ \hline			
			\begin{tabular}[c]{@{}l@{}} AT$^{2}$ \end{tabular}  &\multicolumn{1}{c|}{0.842} & 0.885  &\multicolumn{1}{c|}{0.813} & 0.862  &\multicolumn{1}{c|}{0.805} & 0.853  &\multicolumn{1}{c|}{0.747} & 0.803  \\ \hline
			\begin{tabular}[c]{@{}l@{}} AT$^{3}$ \end{tabular}&\multicolumn{1}{c|}{0.838} & 0.888  &\multicolumn{1}{c|}{0.814} & 0.865  &\multicolumn{1}{c|}{0.802} & 0.857  &\multicolumn{1}{c|}{0.803} & 0.862 \\\clineB{1-9}{3}
			\begin{tabular}[c]{@{}l@{}} wRA \end{tabular}&\multicolumn{1}{c|}{0.951} & 0.959  &\multicolumn{1}{c|}{0.940} & 0.944  &\multicolumn{1}{c|}{0.929} & 0.943  &\multicolumn{1}{c|}{0.906} & 0.923 \\ \hline
			\begin{tabular}[c]{@{}l@{}} wAT$^{1}$ \end{tabular}   &\multicolumn{1}{c|}{0.952} & 0.961  &\multicolumn{1}{c|}{0.945} & 0.954  &\multicolumn{1}{c|}{0.922} & 0.941  &\multicolumn{1}{c|}{0.909} & 0.931 \\ \hline
			\begin{tabular}[c]{@{}l@{}} wAT$^{2}$ \end{tabular}  &\multicolumn{1}{c|}{0.830} & 0.881  &\multicolumn{1}{c|}{0.802} & 0.849  &\multicolumn{1}{c|}{0.814} & 0.859  &\multicolumn{1}{c|}{0.814} & 0.868  \\ \hline
			\begin{tabular}[c]{@{}l@{}} wAT$^{3}$ \end{tabular}&\multicolumn{1}{c|}{0.814} & 0.862  &\multicolumn{1}{c|}{0.779} & 0.828  &\multicolumn{1}{c|}{0.797} & 0.836  &\multicolumn{1}{c|}{0.708} & 0.765  \\\clineB{1-9}{3}
			\begin{tabular}[c]{@{}l@{}} mix$^1$ \end{tabular}  &\multicolumn{1}{c|}{0.945} & 0.956  &\multicolumn{1}{c|}{0.928} & 0.943  &\multicolumn{1}{c|}{0.903} & 0.925  &\multicolumn{1}{c|}{0.916} & 0.937  \\ \hline
			\begin{tabular}[c]{@{}l@{}} mix$^2$ \end{tabular}
			&\multicolumn{1}{c|}{0.954} & 0.964  &\multicolumn{1}{c|}{0.940} & 0.952  &\multicolumn{1}{c|}{0.933} & 0.948  &\multicolumn{1}{c|}{0.912} & 0.934  \\\clineB{1-9}{3}
	\end{tabular}
\end{table*}

Now, when we compare Eq.~(\ref{AT1a}) with the definition of the RA index~(\ref{RA}), we see that they are very similar. In particular, in both indices, RA and AT$^1$, each neighbor of the pair $(i,j)$ is weighted inversely proportional to its degree. The difference between the two measures is that our AT$^1$ index additionally takes into account the entire triad's neighborhood, which is represented by nodes adjacent to the edges $(z,i)$ and $(z,j)$. Comparing the effectiveness of both measures (see Fig.~\ref{fig6} and Tab.~III), there is a slight argument in favour of the AT$^1$ index, which proves the legitimacy of the presented reasoning. In the same vein, and for the sake of completeness, it is worth noting that our weighed wAT$^1$ index is almost identical to the RA weighted index \cite{2010EPLLu,2016SciRepZhu}: 
\begin{eqnarray}\label{wRA}
	\mbox{wRA}_{\,ij}&=&\sum_{z\in\Gamma(i)\cap\Gamma(j)}\frac{w_{zi}+w_{zj}}{s_{z}},
\end{eqnarray}
although our wAT$^1$ score performs slightly better than wRA (see~Tab.~III).

To complete the discussion on the role of the asymmetry of social interactions in link predictability, we also examined other similarity scores accounting for the concept of 'asymmetric triad' (see Fig.~\ref{fig5}). More precisely, we tested the following 'structural' measures:
\begin{eqnarray}\label{AT2}
	\mbox{AT}^2_{\,ij}&=&\sum_{z\in\Gamma(i)\cap\Gamma(j)}(Q_{iz}+Q_{jz}), \\\label{AT3}\mbox{AT}^3_{\,ij}&=&\sum_{z\in\Gamma(i)\cap\Gamma(j)}(Q_{iz}+Q_{zj}),
\end{eqnarray}
and their 'weighted' counterparts:
\begin{eqnarray}\label{wAT2}
	\mbox{wAT}^2_{\,ij}&=&\sum_{z\in\Gamma(i)\cap\Gamma(j)}(v_{iz}+v_{jz}), \\\label{wAT3}\mbox{wAT}^3_{\,ij}&=&\sum_{z\in\Gamma(i)\cap\Gamma(j)}(v_{iz}+v_{zj}),
\end{eqnarray}
and found that they give significantly worse results (see Fig.~\ref{fig6} and Tab.~III). This observation clearly confirms the general message of our paper that the asymmetry of social interactions is important, even in networks where the edges are not formally assigned a direction.

At this point, we would like to refer to the RA score once again. The original rationale for this index is related to the efficiency of \textit{resource transmission between the nodes $i$ and $j$, through their nearest neighbours} \cite{2009EPJBZhou}. There is, however, a subtle inaccuracy between the above reasoning and the RA definition (\ref{RA}). Namely, the reasoning seems to apply to the directed triad T$_{3}$ shown in Fig.~\ref{fig5} and quantified by our AT$^3$ index rather than by the RA-like, our AT$^1$ index that describes the T$_1$ triad. Correspondingly, the success of the measure AT$^1$ (and also wAT$^1$) can be justified as follows: The asymmetric overlap $Q_{zi}$ acts as a proxy for the amount of resources the node $z$ invests in collaborating with the node $i$. If two such investments, e.g. $Q_{zi}$ and $Q_{zj}$, are resource-intensive, they probably cannot be independent, simply because personal resources of $z$ are limited (e.g. there are only 24 hours in a day). This observation stands in line with the study by Dunbar \cite{1992Dunbar}, who concluded that humans, even very active, have a limited capacity to maintain significant interpersonal relationships. In the case of the collaboration network such a dependency between both $z$’s investments simply means a joint project or publication co-authored by $z$, $i$ and $j$, so the link between $i$ and $j$ has to be expected. Please note that such a dependency cannot be simply derived from investments $Q_{iz}$ and $Q_{jz}$, or $Q_{iz}$ and $Q_{zj}$ standing behind the indices AT$^2$ or AT$^3$, respectively. Please also note that such a perspective of resource allocation is completely different from the original one, which we have already questioned.

\subsubsection{Towards intrinsically asymmetric similarity scores}

Finally, while it was not our intention in this paper to search for better than already existing similarity scores for link prediction, nor to argue that the local network measures we introduced, such as asymmetric neighborhood overlap, are generally better to design such indices, we cannot pass over the fact that through trial and error, we have managed to find indices whose effectiveness exceeds (at least in the analysed networks of scientific collaboration) the effectiveness of all others that were examined in this paper. These measures are defined as linear combinations of the previous ones (see Fig.~\ref{fig6} and Tab.~III):
\begin{equation} 
	\mbox{mix}^1_{\,ij}=\mbox{wAT}^1_{ij}+\mbox{QQ}_{\,ij},
\end{equation}
and
\begin{equation} 
	\mbox{mix}^2_{\,ij}=\mbox{wAT}^1_{ij}+\mbox{QA}_{\,ij},
\end{equation}
where QQ, QA and wAT$^1$ are given by Eqs.~(\ref{QQ}), (\ref{QA}) and (\ref{wAT1}), respectively.

\section{Concluding remarks}\label{theend}

The leitmotif of this paper is the problem of "relativity" (or the lack of symmetry) in social relations. To draw attention to this problem, we focused on coauthorship networks, and used the known controversy regarding their atypical weight-topology correlations to show that taking the asymmetry into account can change the understanding of even well-established findings, such as that scientific collaboration networks do not satisfy the Granovetter's strength of weak ties hypothesis. 

More precisely, in this paper, by analysing three different real coauthorship networks (DBLP, APS, and CondMat) and their reliable synthetic model, we show that the networks show strong positive correlations between tie strength, $v$, and neighbourhood overlap, $Q$, of the connected nodes only when both measures take into account the lack of symmetry of the relationship. The observed correlations satisfy the power law scaling: $Q\propto v^\beta$, with the same characteristic exponent $\beta\simeq0.65$ for all studied networks.

In light of the noticed strong correlations, research on link prediction methods that would take advantage of link asymmetry seems particularly interesting. By testing various link scores used in similarity-based unsupervised link and weight prediction methods \cite{2015SciRepZhao, 2016SciRepZhu, 2020aIEEELi, 2020bIEEELi}, we argue that taking into account the asymmetry of social ties can remarkably increase efficiency of these methods. We are also convinced that taking into account the asymmetry of social ties can also improve more advanced prediction methods, especially those supervised \cite{2018IEEEChenbo}. Finally, since in many ways, scientific collaboration networks are very specific, a natural continuation of the research presented here would be to check whether similar results can be obtained by analysing other (more typical) social networks, or even other complex networks, not necessarily social ones.

\section*{Acknowledgments} 

This research was funded by the POB Research Centre Cybersecurity and Data Science of Warsaw University of Technology within the Excellence Initiative Program - Research University (ID-UB).

\section*{Supplementary Matherials} 

All the data used in this paper as well as the software developed for calculation of overlaps and for prediction can be found at \cite{SM}.

\bibliography{granovetter22}

\end{document}